\documentclass{aastex}
\usepackage{emulateapj5}

\def\be{\begin{equation}}
\def\ee{\end{equation}}
\def\ba{\begin{eqnarray}}
\def\ea{\end{eqnarray}}
\def\hatl{{\bf\hat l}}
\def\bB{{\bf B}}

\newenvironment{inlinefigure}{
\medskip
\def\@captype{figure}
\noindent\begin{minipage}{0.999\linewidth}\begin{center}}
{\end{center}\end{minipage}\medskip}

\begin{document}

\title{Warping of Accretion Disks with Magnetically Driven Outflows:\\
A Possible Origin for Jet Precession}
\author{Dong Lai}
\affil{Center for Radiophysics and Space Research, Department of Astronomy, 
Cornell University, Ithaca, NY 14853\\
Email: dong@astro.cornell.edu}

%%%%%%%%%%%%%%%%%%%%%%%%%%%%%%%%%%%%%%%%%%%%%%%%%%%%%%%%%%%%%%%%%
\begin{abstract}
Current theoretical models for the outflows/jets from AGN,
X-ray binaries and young stellar objects involve large-scale magnetic
fields threading an underlying accretion disk. We suggest that 
such a disk is subjected to warping instability
and retrograde precession driven by magnetic torques associated with the 
outflow. The growth timescale for the disk warp and the precession
period are of order the radial infall time of the disk. 
These effects may be relevant to jet precession and other variabilities
observed in many systems.
\end{abstract}

%%%%%%%%%%%%%%%%%%%%%%%%%%%%%%%%%%%%%%%%%%%%%%%%%%%%%%%%%%%%%%%%%
\section{Introduction}

Collimated outflows/jets are produced from active galactic nuclei, 
compact stars in X-ray binaries, and young stellar objects (YSOs) 
(e.g., Krolik 1999; Mirabel \& Rodriguez 1999; 
Fender 2003; Reipurth \& Bally 2001).
%Bridle \& Perley 1984;
Over the last two decades, evidence for jet precession in these systems
has steadily increased. The clearest example is the galactic 
source SS 433, whose jet direction varies with
an approximate 164 day period (Margon 1984; Eikenberry et al.~2001).
% (half angle of the precession cone is 20 degree). 
The black hole binary GRO J1655-40 shows jet precession with a 
period of 3 days (Tingay et al.~1995; Hjellming \& Rupen 1995).
% at an angle of 2 degrees.
The jets from the super soft source CAL83 may be precessing with 
a period of $\sim 69$ days (Cowley et al.~1998).
In AGNs, indirect evidence for jet precession is seen in the morphology 
of the radio hot spots, which show ``fossil'' components offset from
the ``present'' component positions (e.g., Cygnus A). 
%The jets in radio galaxies often appear not to be perpendicular to the 
%dust disks (e.g. Schmitt et al.~2002). 
Finally, the changes in the flow directions of 
several YSO jets have been interpreted in terms of jet precession
(e.g. 
%Mundt \& Eisl\"offel 1998; 
Terquem et al.~1999; Bates et al.~2000).

A natural cause for jet precession is the changes of orientation of
the underlying accretion disk. In addition, the super-orbital 
variabilities observed in a number of X-ray binaries (e.g., the
35-day periodicity in Her X-1; see Priedhorsky \& Holt 1987; 
Ogilvie \& Dubus 2001) have long been interpreted as due to 
precession of a tilted accretion disk. In both X-ray binaries and YSOs, 
the tidal force from the companion star could play a role 
in driving the precession (e.g., Katz 1973; Wijers \& Pringle 1999; 
Terquem et al.~1999; Bate et al.~2000; Ogilvie \& Dubus 2001), although 
it may not be the only or the dominant effect 
(e.g., the 3 day precession period of the GRO J1655-40 jet is too 
short to be explained by the tidal effect for a 2.6 day binary). 
Most importantly, for the precession to operate,
disk tilt needs to be excited and maintained. In accreting 
binary systems, the disk plane is expected to be aligned with 
the orbital plane since the disk angular momentum originates 
from the binary motion. For YSOs in binaries, 
the disk plane may be initially misaligned with the orbital plane. However, 
if we consider the disk as a collection of circular rings, different rings 
will have different precession rates; it has been recognized that the combined
effects of differential precession and internal disk stress/dissipation 
tend to damp the the disk tilt, so that the disk settles into the binary 
plane (Lubow \& Ogilvie 2000). 

Several driving mechanisms for disk tilt/warp have been proposed.
Schandl \& Meyer (1994) showed (in the context of Her X-1) 
that irradiation-driven wind from the outer parts of the disk can carry away
momentum flux and make the disk unstable to warping.
Pringle (1996) showed that even without wind loss, radiation pressure itself
can induce warping instability in the outer region of the disk.
Futher studies indicated this radiation-driven warping instability 
may indeed operate in X-ray binaries (e.g., 
%Maloney \& Begelman 1997;
Maloney, Begelman \& Nowak 1998; Wijers \& Pringle 1999), 
although it does not provide a generic explanation for the long-term 
variabilities in all X-ray binaries (Ogilvie \& Dubus 2001).
Quillen (2001) showed that a wind passing over the disk surface may 
induce warping via Kelvin-Helmholtz instability.
Finally, in the case of disk accretion onto magnetic stars (e.g., neutron
stars, white dwarfs and T Tauri stars), the stellar magnetic field can 
induce disk warping and precession (Lai 1999; see also 
Aly 1980; Lipunov \& Shakura 1980; Terquem \& Papaloizou 2000);
this may explain several observed
features of quasi-periodic oscillations in low-mass X-ray binaries 
(Shirakawa \& Lai 2002a), milli-Hertz variabilities in accreting
X-ray pulsars (Shirakawa \& Lai 2002b), and variabilities of T Tauri
stars (Terquem \& Papaloizou 2000; see also Agapitou et al.~1997).

In this paper we suggest a new disk warping mechanism that is directly 
tied to the production of magnetically driven outflows/jets.
Using an idealized setup (\S 2), we show that 
a disk threaded by a large-scale magnetic field may experience a warping
instability and precess around the central object (\S3). 
These magnetically driven disk warping and precession
arise from the interaction between the large-scale magnetic field and the
induced electric current in the disk. While more studies are needed,
we suggest that these effects may provide a natural explanation for the 
procession of jets/outflows and other variabilities observed 
in various systems (\S 4).  

\section{The Setup}

The current paradigm for the origin of astrophysical jets/outflows involves 
a large-scale magnetic field threading the accretion disk 
around a central object (star or black hole);  this ordered
magnetic field plays a crucial role in extracting/channeling mass, 
energy and angular momentum from the disk. The energy outflow can be either 
hydromagnetic (with significant mass flux) (Blandford \& Payne 1982)
%Lovelace et al.~1986; K\"onigl \& Pudritz 2000) 
or electromagnetic (dominated by Poynting flux) (Blandford 1976; 
Lovelace 1976). The origin of the disk-threading magnetic field 
is not completely clear: the field could be advected inwards by accretion,
or generated locally by dynamo processes.  
In the case of protostellar outflows, the stellar magnetic field may play an
important role (Shu et al.~1994,~2000). 
Many theoretical/numerical studies have been 
devoted to understanding magnetically driven outflows/jets from accretion 
disks (e.g., see recent reviews by Lovelace et al.~1999; 
K\"onigl \& Pudritz 2000; Meier et al.~2001).

\begin{inlinefigure}
\scalebox{.9}{\rotatebox{0}{\plotone{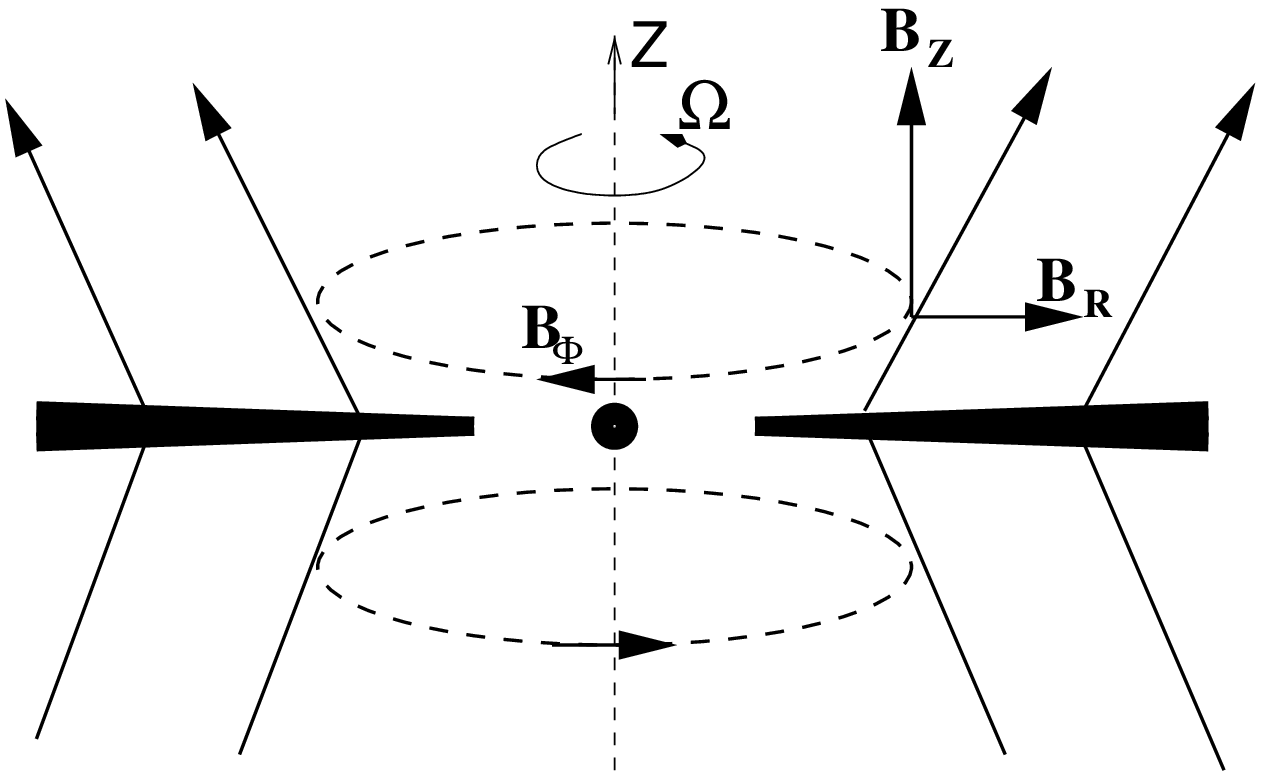}}}
\figcaption{A sketch of the idealized disk -- magnetic field 
configuration in which hydromagnetic outflows are produced. 
The disk is threaded by a large-scale poloidal field, which 
gets twisted by the disk, generating the toroidal field.
}\end{inlinefigure}

Figure 1 shows an idealized sketch of the magnetic field -- disk configuration
in which hydromagnetic outflows are produced. A geometrically 
thin disk is threaded by the poloidal magnetic field 
$\bB_p=B_R{\bf\hat R}+B_Z{\bf\hat Z}$
(where ${\bf\hat R},~{\bf\hat Z}$ are unit vectors in cylindrical coordinates).
This poloidal field is of course curved (with $B_R=0$ at the disk midplane), 
and we use the superscript ``+'' to denote the field at the upper disk 
surface. For a sufficiently inclined field (radial pitch angle
$\theta=\tan^{-1}|B_R^+/B_Z|>30^\circ$), centrifugally driven outflows 
are possible (Blandford \& Payne 1982).  
The twisting of $B_p$ by the disk and the outflow gives rise to
a toroidal field $B_\Phi$, which has different signs
above and below the disk plane. We introduce the azimuthal pitch
$\zeta$, so that $B_{\Phi}^+=-\zeta B_Z$ at the upper disk surface.
That $\zeta$ is positive reflects the fact that the disk always rotates
faster than the outflow\footnote{In the stationary, axisymmetric, MHD 
wind theory, $B_\Phi/B_p=R(\Omega-\Omega_d)/V_p$, where $V_p(R,Z)$ is the 
the poloidal velocity of the outflow and $\Omega(R,Z)$ its rotation
rate, $\Omega_d$ is the disk rotation at the midplane. Alternatively,
in the absence of outflow, $B_\Phi$ near the disk is governed
schematically by the equation $\partial B_\Phi/\partial t=RB_Z
{\partial\Omega/\partial Z}-B_\Phi/\tau_{\rm diss}$, where 
$\tau_{\rm diss}$ is the disspation timescale. In steady state, we find
$B_\Phi^\pm=-\zeta B_z$, with $\zeta=(R/H_B)\Omega\tau_{\rm diss}$
(where $H_B$ is the vertical scale height in which $\Omega$ in the 
magnetosphere varies.}.  
The discontinuities in $B_\Phi$ and $B_R$ across the disk 
imply a net disk surface current (integrated over the disk thickness)
${\bf K}=(c/2\pi)(-B_\Phi^+{\bf\hat R} + B_R^+ {\bf\hat\Phi})$.
Thus, for an unpertutbed (unwarped) disk, the only torque on the disk 
(per unit area) is the usual magnetic braking torque
\be
T_Z={1\over 2\pi}\,R\,B_Z\,B_{\Phi}^+=-{1\over 2\pi}\,\zeta\,R\,B_Z^2.
\label{eqtz}\ee
Angular momentum is extracted out of the disk by the open field lines,
and can be carried away by torsional Alfven waves
or a centrifugally driven wind (for sufficiently inclined
poloidal fields). In addition, the radial magnetic force,
\be 
F_R=B_ZB_R^+/(2\pi),
\label{eqfr}\ee
provides partial support of the disk 
against gravity.

The configuration shown in Fig.~1 lies at the heart of many 
theoretical models of magnetic disk outflows. The real situation is 
certainly much more complicated. Nevertheless, in \S 3 we shall adopt 
this simple picture in our consideration of magnetically driven
disk warping and precession --- our goal is to discuss these physical effects
in a way that is as transparent as possible, so as to expose their 
robustness and potential pitfall.

%%%%%%%%%%%%%%%%%%%%%%%%%%%%%%%%%%%%%%%%%%%%%%%%%%%%%%%%%%%%%%%%%
\section{Magnetically Driven Disk Warping and Precession}

We now consider warping perturbations to the disk. 
The disk can be considered as a collection of circular 
rings which interact with each other via internal stresses. We specify 
the disk warping by the unit normal vector ${\bf\hat l}(r)$
for the ring at radius $r$. Consider the situation (see Fig.~2)
in which the ring around radius $r$ is tilted by an angle $\beta$ 
while most of the disk remains flat ($\hatl_1=\hatl_2={\bf\hat Z}$, 
where ${\bf\hat Z}$ is the normal vector of the unperturbed disk; see Fig.~1). 
Since the magnetic field $B_Z$ is produced by all the currents in the 
disk-outflow system, we posit that $B_Z$ at radius $r$ is not modified 
by such a local perturbation. For the ring at $r$, the field ${\bf B}_Z$ 
acts as an ``external'' magnetic field. Now set up a coordinate 
system with the $z$-axis along the $\hatl(r)$ 
(see the lower panel of Fig.~2) and the associated cylindrical coordinate
system $(r\phi z)$. As in the case of the unperturbed disk,
twisting of $B_z=B_Z\cos\beta$ by the disk rotation produces a toroidal field
(in the new coordinate system) $B_{\phi,t}$ (where the subscript ``t'' 
serves as a reminder that this field is due to twisting of $B_z$),
whose value (at radius $r$) at the upper disk plane is $B_{\phi,t}^+
=-\zeta B_z$, and at the lower disk plane $B_{\phi,t}^-=\zeta B_z$.
Note that $B_z$ is necessarily continuous across the disk plane 
in this perturbation. The radial surface current associated with 
$(B_{\phi,t}^+ - B_{\phi,t}^-)$ is
\be
K_r=-{c\over 2\pi} B_{\phi,t}^+={c\over 2\pi}\zeta B_Z\cos\beta.
\ee
Interaction of this $K_r$ with $B_z$ gives an azimuthal force
on the disk, and results in the magnetic braking torque (per unit area)
$T_z=-\zeta r B_z^2/(2\pi)=-\zeta r B_Z^2\cos^2\beta/(2\pi)$,
analogous to eq.~(\ref{eqtz}). Unlike the unperturbed disk, however,
the ``extrenal'' field $\bB_Z$ now has a $\phi$-component, $B_\phi=B_Z\sin\beta
\cos\phi$, and the interaction of $K_r$ with $B_\phi$ gives rise to a vertical
force (per unit area)
\be
F_z={1\over 2\pi}\,\zeta\,B_Z^2\sin\beta\,\cos\beta\,\cos\phi.
\ee
This vertical force can also be seen as arising from the imbalance 
of magnetic stresses above and below the disk plane: 
$B_{\phi}^+=B_\phi+ B_{\phi,t}^+$ in the upper disk plane 
while $B_{\phi}^-=B_\phi- B_{\phi,t}^+$ in the lower disk plane, 
and thus $F_z=[(B_{\phi}^-)^2-(B_{\phi}^+)^2]/(8\pi)$. Obviously,
averaging the force over the azimuthal angle, $\langle F_z\rangle=(1/2\pi)
\int_0^{2\pi}F_z\,d\phi$, gives zero. But since $F_z$ is unevenly distributed
along the ring, it gives rise to a net torque on the ring. 
Averaging over $\phi$, the torque per unit area is
\ba
\langle {\bf T}_{\rm warp}\rangle &=& 
-{1\over 4\pi}\,\zeta\,r B_Z^2\cos\beta\sin\beta
\,{\bf\hat y}\nonumber \\
&=& -{1\over 4\pi}\,\zeta\,r B_Z^2\cos\beta \,
\left[{\bf\hat Z}-({\bf\hat Z}\cdot\hatl)\right].
\ea
Clearly, this torque tends to pull the vector $\hatl(r)$ away from 
``external'' magnetic field direction ${\hat{\bf B}_Z}$. 
This is the {\it magnetically driven warping
instability}: for small tilt, the angle $\beta$ tends to grow (in the 
absence of other forces) at the rate
\be
\Gamma_{\rm warp}={\zeta B_Z^2\over 4\pi \Sigma\,r\,\Omega_d},
\label{gwarp}\ee
where $\Sigma$ is the surface density, and $\Omega_d$ 
the angular velocity of the disk.
In a hypothetical situation, if ${\bf B}_Z$ had a fixed direction for all
$\beta$, and if there were no coupling between different rings, the vector
$\hatl(r)$ would evolve toward being perpendicular to ${\bf B}_Z$,
i.e., the plane of the disk (ring) would prefer to lie along the ``external''
magnetic field (cf. Lai 1999). 

\begin{inlinefigure}
\scalebox{.9}{\rotatebox{0}{\plotone{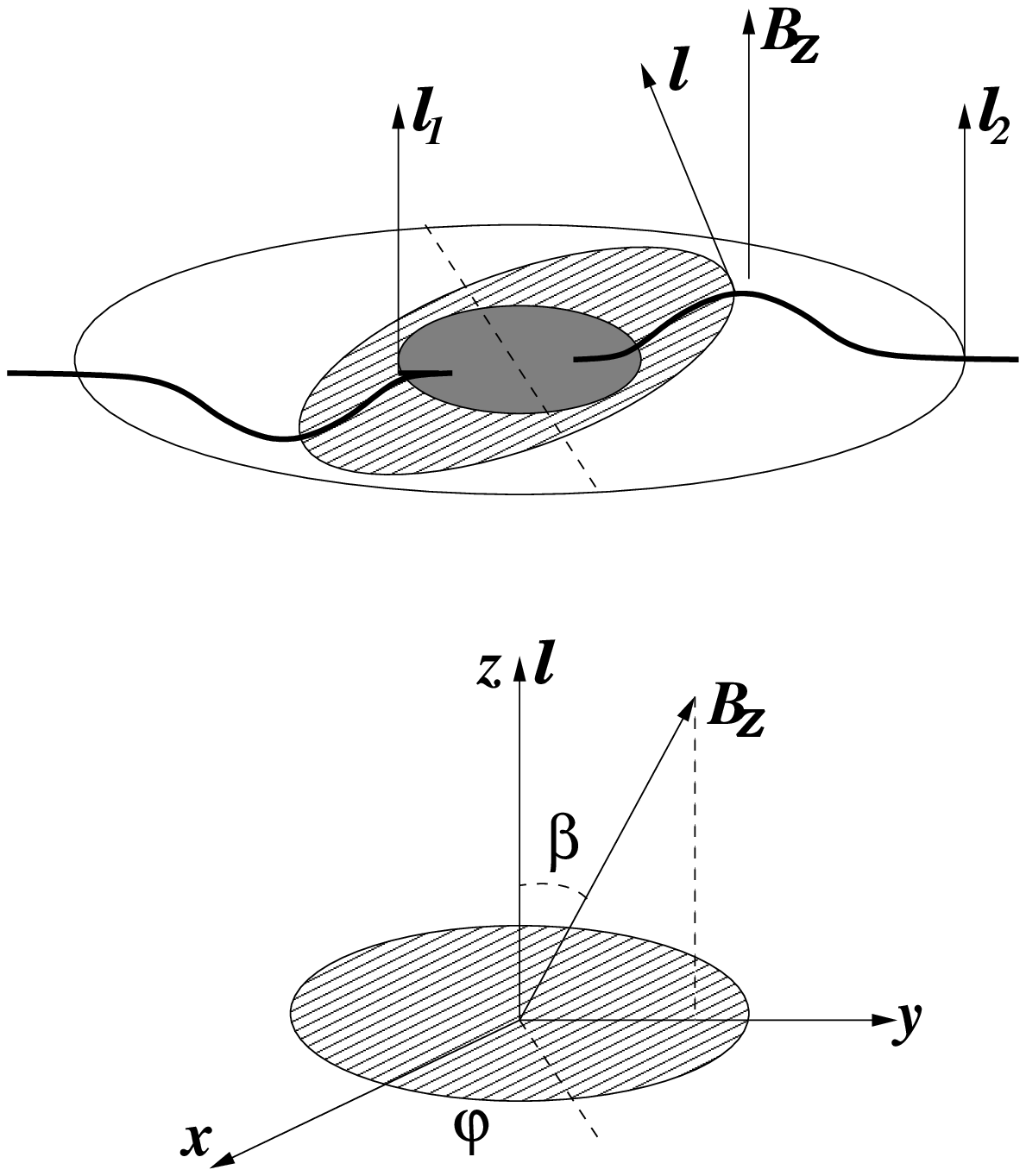}}}
\figcaption{Warping perturbation to the disk around radius
$r$. For clarity, only the $Z$-component of the magnetic field,
$\bB_Z$, is shown.
}\end{inlinefigure}

In the perturbed disk, the azimuthal surface current $K_\phi
\simeq K_\Phi=(c/2\pi)B_{R}^+=(c/2\pi)B_Z\tan\theta$ (for small
$\beta$) also interacts
with the ``external'' field $\bB_Z$, which can be decomposed into 
$\bB_Z=B_Z\cos\beta\,{\bf\hat z}+B_Z\sin\beta\,(\sin\phi
\,{\bf \hat r}+\cos\phi\,{\bf\hat\phi})$.
The interaction between $K_\phi$ and $B_z=B_Z\cos\beta$ 
gives a radial force, as in eq.~(\ref{eqfr}).
For the perturbed disk ($\beta\ne 0$), a new vertical force
arises from the interaction between $K_\phi$ and 
$B_r=B_Z\sin\beta\sin\phi$:
\be
F_z=-{1\over c}\,K_\phi B_Z\sin\beta\,\sin\phi.
\ee
This force is also equal to $[(B_{r}^-)^2-(B_{r}^+)^2]/(8\pi)$, where
$B_{r}^\pm=B_Z\sin\beta\sin\phi\pm B_Z\tan\theta$.
Again, $\langle F_z\rangle=0$, but the net torque is non-zero.
Averaging over $\phi$, we find the torque per unit area
\be
\langle {\bf T}_{\rm prec}\rangle
=-{1\over 2c}\,K_\phi B_z r\,\sin\beta\,{\bf\hat x}
=-{1\over 2c}\,K_\phi B_z r\,{\bf\hat{Z}}\times\hatl.
\ee
Clearly, the effect of this torque is to make the ring (at radius $r$)
precess around the $Z$-axis (the unperturbed disk normal vector).
The angular frequnecy of the {\it magnetically driven precession} is
\be
{\bf \Omega}_{\rm prec}=-{K_\phi B_Z\over 2\,c\,\Sigma\,r\,\Omega_d}
\,{\bf\hat{Z}}=-{B_Z^2\tan\theta\over 4\pi\,\Sigma\,r\,\Omega_d}
\,{\bf\hat{Z}}.
\label{oprec}\ee
The negative sign implies that the pecession is retrograde,
i.e., opposite to the disk rotation.

%%%%%%%%%%%%%%%%%%%%%%%%%%%%%%%%%%%%%%%
\section{Discussion}

To get an idea on the warping and precession timescales 
associated with eqs.~(\ref{gwarp}) and (\ref{oprec}), 
we note the disk angular momentum
equation (for an unwarped disks) can be written as 
\be
\Sigma V_R{d\over dR}(\Omega_d R^2)=-{1\over 2\pi}\zeta RB_Z^2+T_{\rm visc},
\ee
where $V_R$ is the radial velocity and $T_{\rm visc}$ is the 
viscous torque. If we set $T_{\rm visc}=(f-1)(-\zeta RB_Z^2/2\pi)$ and
use $\Omega_d\simeq (GM/R^3)^{1/2}$, we find
\be 
\Gamma_{\rm warp}={\zeta\over \tan\theta}|\Omega_{\rm prec}|
=-{V_R\over 4fR}.
\ee
For typical parameters ($\zeta\sim\tan\theta\sim f\sim 1$),
the warping timescale and precession period are of order the disk infall 
time $R/|V_R|$. 

It is important to note that the magnetically driven disk warping
and precession discussed in \S 3 are secular effects which operate
on timescales much longer than the disk dynamical time.
In deriving eqs.~(\ref{gwarp}) and (\ref{oprec}), we have
implicitly assumed that the perturbed disk (see Fig.~2) reaches a
new steady state so that the magnetic field behaves
in a similar way as in the unperturbed disk (e.g.,
$B_{\phi,t}^\pm=\mp \zeta B_z$). These effects are not captured in a dynamical
perturbation analysis of the disk. 
Clearly, our derivation of the magnetically driven
warping and precession is far from being rigorous. Nevertheless, it suggests 
that the warped, precessing disk-outflow may be an alternative (and preferred)
state for the accretion disk threaded by large-scale magnetic fields.

The effects of magnetically driven warping and precession
discussed in this paper are analogous to 
the similar effects that exist in the accretion disks around magnetic 
stars (Lai 1999; Shirakawa \& Lai 2002a,b; Pfeiffer \& Lai 2003;
see also Aly 1980; Lipunov \& Shakura 1980). In those systems,
the central star provides the ``external'' magnetic field for the disk.
The warping and precessional torques arise from the interaction 
between the surface current on the disk and the horizontal magnetic 
field (parallel to the disk) produced by the stellar magnetic dipole. 
Here, in the absence of the central magnetic star, the ``external''
magnetic field (for the warped region of the disk) 
is generated by the other currents in the disk-outflow system,
and we have argued that similar torques exist:
the warping torque relies on the surface current
generated by the twisting of the vertical field threading the disk, while the 
precessional torque relies on the azimuthal screening current
due to the diamagnetic response of the disk. 
The simplistic nature of our arguments (perhaps
even to the point of being wrong) in \S 3 and the complexity of
the real astrophysical systems (see references in \S 2) preclude us from
addressing many important issues such as the extent and location of the
warped disk region. Nevertheless, we are not aware of 
any previous discussion of these effects, and we think it is useful to
present them so that they may be studied further by the astrophysics 
community.

The magnetically driven warping and precession effects discussed in this 
paper potentially have advantages over the other disk warping 
mechanisms that are needed to explain jet precession (see \S 1). 
For example, the radiation driven warping instability (Pringle 1996),
while probably important for X-ray binaries (Ogilvie \& Dubus 2001),
is irrelevant for YSOs (since the instability operates
at large disk radii). The magnetic effects discussed in this paper 
are directly connected to the large-scale magnetic field associated with the
jet/outflow, and the model naturally predicts retrograde precession 
of the warped disk (and thus the jet), in agreement with observations
in X-ray binaries (e.g., Her X-1).

\acknowledgments
The idea for this paper was motivated by the questions from
Chris McKee and Frank Shu following a seminar that the author
gave at UC Berkeley in 2001. I thank Richard Lovelace for
useful discussion. This work is supported in part by NSF Grant AST 9986740 and
NASA grant NAG 5-12034, as well as by a research fellowship 
from the Alfred P. Sloan foundation.

%%%%%%%%%%%%%%%%%%%%%%%%%%%%%%%%%%%%%%%%%%%%%%%%%%%%%%%%%%%%%%%%%%%%

%\clearpage
%\begin{figure}
%\plotone{f1.eps}
%\caption{A sketch of the idealized disk -- magnetic field 
%configuration in which hydromagnetic outflows are produced. 
%The disk is threaded by a large-scale poloidal field, which 
%gets twisted by the disk, generating the toroidal field.
%}\end{figure}

%\clearpage
%\begin{figure}
%\plotone{f2.eps}
%\caption{Warping perturbation to the disk around radius
%$r$. For clarity, only the $Z$-component of the magnetic field,
%$\bB_Z$, is shown.
%}\end{figure}

\end{document}